\documentclass[british,sort&compress]{iopart}
\usepackage[T1]{fontenc}
\usepackage[latin9]{inputenc}
\usepackage{geometry}
\usepackage{xcolor}
\usepackage{babel}
\usepackage{graphicx}
\usepackage{esint}
\usepackage[numbers]{natbib}
\PassOptionsToPackage{normalem}{ulem}
\usepackage{ulem}
\usepackage[unicode=true,pdfusetitle,
 bookmarks=true,bookmarksnumbered=true,bookmarksopen=true,bookmarksopenlevel=1,
 breaklinks=true,pdfborder={0 0 1},backref=false,colorlinks=false]
 {hyperref}

\makeatletter

\providecolor{changesadded}{rgb}{0.26,0.44,0.68}
\providecolor{changesdeleted}{rgb}{1,0.64,0.28}

\DeclareRobustCommand{\lyxsout}[1]{\ifx\\#1\else\sout{#1}\fi}

\usepackage{iopams}
\usepackage{setstack}

\usepackage{marginnote}

\newcommand{\eqref}[1]{(\ref{#1})}

\makeatother

\begin{document}
\title[Near-field enhancement by waveguide-plasmon polaritons in ...]{Near-field enhancement by waveguide-plasmon polaritons in a nonlocal
metasurface}
\author{Xiaorun Zang and Andriy Shevchenko}
\address{Department of Applied Physics, Aalto University, P.O. Box 13500, FI-00076
Aalto, Finland}
\ead{xiaorun.zang@aalto.fi}
\begin{abstract}
Localized surface plasmons in metal nanoparticles are widely used
in nano-optics to confine and enhance optical fields. It has been
previously shown that, if the nanoparticles are distributed periodically,
an additional enhancement can be achieved by coupling the localized
surface plasmons to the diffraction orders of the lattice, forming
surface lattice resonances. In this work, we study an even further
improvement of the near-field enhancement by placing a metal-dielectric
slab waveguide beneath the lattice of the particles to excite coupled
waveguide-plasmon polaritons. These excitations can extend over many
periods of the lattice, making the metasurface highly nonlocal. We
numerically demonstrate that the approach can provide a significant
extra increase in the near-field intensity -- by a factor of 80 over
that produced by a single-particle plasmon resonance and by 7 over
the lattice-resonance enhancement. The described enhancement mechanism
can be used to design extraordinarily efficient nonlocal optical metasurfaces
for many applications, including surface-enhanced Raman spectroscopy,
fluorescence spectroscopy, nonlinear optics, and solar energy harvesting.
\end{abstract}
\noindent{\it Keywords\/}: {Metasurfaces, Waveguide-plasmon polaritons, Surface plasmons, Bloch
wave, SERS, Plasmonics, Field enhancement}
\maketitle

\section{Introduction}

Local enhancement of optical fields plays an essential role in many
applications of optics and photonics, including optical sensors and
detectors \citep{liedbergSurfacePlasmonResonance1983,homolaSurfacePlasmonResonance1999},
light sources \citep{oultonPlasmonLasersDeep2009,oultonSurfacePlasmonLasers2012,liuIntegratedNanocavityPlasmon2016,liangPlasmonicNanolasersOnchip2020},
photovoltaic devices \citep{schaadtEnhancedSemiconductorOptical2005,pillaiPlasmonicsPhotovoltaicApplications2010,atwaterPlasmonicsImprovedPhotovoltaic2010},
and nonlinear optical components \citep{bouhelierNearfieldSecondharmonicGeneration2003,berthelotSilencingEnhancementSecondharmonic2012,czaplicki*LessMoreEnhancement2018}.
An efficient near-field enhancement can be obtained with the help
of metal nanostructures that exhibit localized surface plasmon resonances
(LSPRs) \citep{klarSurfaceplasmonResonancesSingle1998,maierPlasmonicsFundamentalsApplications2007,ciraciProbingUltimateLimits2012}.
Due to a high density of metal electrons, light can be strongly squeezed
into subwavelength regions \citep{ciraciProbingUltimateLimits2012}.
The squeezing is especially effective near locations of abrupt and
small-volume inhomogeneities of the structures, including interparticle
gaps and sharp edges. At such inhomogeneities, very high field intensities
can be obtained. For example, the optical intensity in the gap of
a gold bowtie antenna consisting of two tip-to-tip facing prisms with
sharp corners provides a higher enhancement than the analogue dimer
of nanodiscs with round surfaces \citep{haoElectromagneticFieldsSilver2004,dodsonOptimizingElectromagneticHotspots2013}.
These \textquotedbl hot spots\textquotedbl{} are in the core of many
applications, for example, in surface-enhanced Raman spectroscopy
(SERS) \citep{kneippSingleMoleculeDetection1997,nieProbingSingleMolecules1997}.
The electromagnetic enhancement factor of the SERS signal is given
by $\eta_{\mathrm{SERS}}\approx|E/E_{0}|^{4}$ \citep{kneippSingleMoleculeDetection1997},
where $E$ and $E_{0}$ are, respectively, the total and incident
electric fields in the hot spot.

Nanoparticles of various materials, compositions, shapes, sizes, and
surrounding dielectric environments have been explored to control
the local field intensity \citep{haesPlasmonicMaterialsSurfaceenhanced2005,xiaShapecontrolledSynthesisSurface2005,ahmadivandToroidalMetaphotonicsMetadevices2020}.
In addition to conventional approaches, high-index dielectric nanoparticles
have been recently used to resonantly enhance both the electric and
magnetic fields while significantly reducing the ohmic losses \citep{caldarolaNonplasmonicNanoantennasSurface2015,kuznetsovOpticallyResonantDielectric2016,alessandriEnhancedRamanScattering2016,weitzComparisonRamanScattering1983}.
Plasmonic dimers that exhibit a toroidal dipole resonance have been
shown to outperform the conventional disc dimers regarding the field
enhancement and third harmonic generation \citep{ahmadivandToroidalDipoleenhancedThird2019}.
It has also been shown that, reducing the gap size increases the local
field intensity up to a certain limit dictated by quantum effects,
such as the effect of electron tunnelling through the gap and nonlocal
screening \citep{garciadeabajoNonlocalEffectsPlasmons2008,estebanBridgingQuantumClassical2012,luoSurfacePlasmonsNonlocality2013,zhuQuantumMechanicalLimit2014,girardMolecularDecayRate2015,zhuQuantumMechanicalEffects2016}.
On the other hand, the quantum confinement effect can play a crucial
role for plasmonic particles that are smaller than $10$ nm in size
\citep{garciadeabajoNonlocalEffectsPlasmons2008}. Recently, quantum
confinement of electrons in a one-dimensional metallic single-walled
carbon nanotube with a diameter of $1$-$2$ nm was reported to form
a quantized Luttinger-liquid plasmon that exhibits a strong spatial
confinement and low attenuation loss. Such quantum plasmons could
be useful for plasmonic integrated devices \citep{shiObservationLuttingerliquidPlasmon2015}.

Much higher local fields can be obtained by arranging plasmonic nanoparticles
in a lattice \citep{vecchiShapingFluorescentEmission2009,rodriguezCouplingBrightDark2011}
or superlattice \citep{wangBandedgeEngineeringControlled2017,kravetsPlasmonicSurfaceLattice2018}.
In this case, the field scattered by one particle can arrive at the
neighbouring particles in phase with the other scattered fields, when
the periods are appropriate \citep{wangRichPhotonicWorld2018,cherquiPlasmonicSurfaceLattice2019,kravetsPlasmonicSurfaceLattice2018}.
This results in  collective LSPRs, which have also been called plasmonic
surface lattice resonances \citep{garciadeabajoColloquiumLightScattering2007,kravetsPlasmonicSurfaceLattice2018,lunnemannDispersionGuidedModes2014}
or collective lattice modes \citep{wangRichPhotonicWorld2018,cherquiPlasmonicSurfaceLattice2019,kravetsPlasmonicSurfaceLattice2018,utyushevCollectiveLatticeResonances2021}.
These excitations lead to a further increase in the local field around
each nanoparticle. It has been estimated that a surface lattice resonance
can increase the local field intensity by an order of magnitude \citep{zouSilverNanoparticleArray2004,rossOpticalPropertiesOne2016,lauxSingleEmitterFluorescence2017}.
In the case of a superlattice, the field enhancement can be achieved
simultaneously at multiple modes \citep{wangBandedgeEngineeringControlled2017}.
Moreover, LSPRs can be coupled to optical modes guided in the substrate
of a planar metasurface \citep{dingReviewGapsurfacePlasmon2018},
leading to the so-called waveguide-plasmon polaritons (WPPs) \citep{christWaveguideplasmonPolaritonsStrong2003,rodriguezLightemittingWaveguideplasmonPolaritons2012}.
It has been shown that a hyperbolic metasurface can support such hybrid
waves or quasi-TE and quasi-TM modes \citep{yermakovHybridWavesLocalized2015}.
Very recently, an array of lossy plasmonic nanoparticles on a dielectric
slab waveguide was shown to guide a resonant mode of hybrid polarization
states with an ultrahigh quality factor \citep{kolkowskiEnablingInfiniteFactors2023}.
WPPs hold great potential for yielding extra near-field enhancement
in arrays of plasmonic nanoparticles, and in our opinion, an in-depth
investigation into this mechanism is highly relevant and important.

In this work, we design and study a nonlocal metasurface \citep{overvigDiffractiveNonlocalMetasurfaces2022}
that provides a significant additional near-field enhancement by WPPs.
The metasurface consists of a lattice of silver nanodimers mounted
on a metal-dielectric slab waveguide. Besides a gap enhancement in
each nanodimer, an additional intensity enhancement -- by a factor
of $80$ -- arises from the coupling of LSPRs to the guided Bloch
modes. These modes belong to the second stop band of the structure,
exhibiting a collective resonance at normal incidence. Metasurfaces
of this type can be used for many applications, but for the purpose
of demonstration, we consider a SERS substrate operating at a wavelength
of $780$ nm, one of typical wavelengths in Raman spectroscopy \citep{shevchenkoLargeareaNanostructuredSubstrates2012}.
At this wavelength, the SERS enhancement in each gap of the dimers
is more than $6000$ times higher than that in an isolated dimer on
a semi-infinite dielectric. The metasurface can easily be redesigned
to enhance an optical field at any other wavelength. In addition,
nonlocal metasurfaces consisting of meta-atoms other than nanodisc
dimers, such as bowtie antennae or trimers, are expected to benefit
from the extra near-field enhancement by WPPs as well.

The paper is organized as follows. In Section~\ref{sec:slab}, we
study the guided modes of a one-dimensional slab waveguide. This study
provides a starting point for the design of a planar metasurface and
a physical insight into the properties of possible guided Bloch modes.
In Section~\ref{sec:Bloch_metasurfaces}, we establish a connection
between the near-field enhancement spectra in a designed planar metasurface
and the empty-lattice Bloch-mode band diagrams. In Section~\ref{sec:nf_metasurface},
we optimize the designed metasurface to maximize its near-field enhancement
factor. To achieve a significant additional enhancement of light in
the gap hot spots, we match the frequencies of the lattice resonances
and the guided modes by varying the lattice periods and slab thickness.
We draw conclusions in Section~\ref{sec:Conclusions}.

\section{Guided modes in a slab waveguide\label{sec:slab}}

\begin{figure}[tb]
\begin{centering}
\includegraphics{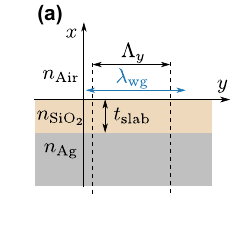}~~~~\includegraphics{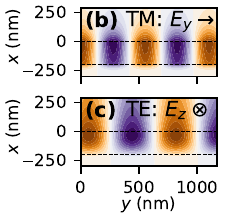}
\par\end{centering}
\begin{centering}
\includegraphics[scale=0.95]{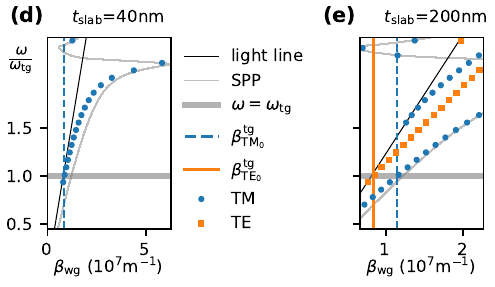}
\par\end{centering}
\centering{}\caption{(a) A schematic diagram of a silver-silica-air slab waveguide. From
top to bottom, the air, silica and silver layers have refractive indices
$n_{\mathrm{air}}$, $n_{\mathrm{SiO_{2}}},$ and $n_{\mathrm{Ag}}$,
respectively. The silica slab layer has a thickness of $t_{\mathrm{slab}}$.
Field profiles of the fundamental (b) TM and (c) TE modes when the
slab thickness is $t_{\mathrm{slab}}=200$ nm. The dispersion curves
are plotted for (d) $t_{\mathrm{slab}}=40$ nm and (e) $t_{\mathrm{slab}}=200$
nm. Blue and orange dots indicate the TM and TE modes, respectively.
At the target frequency with $\omega/\omega_{\mathrm{tg}}=1$, the
propagation constants of the fundamental TM mode $\beta_{\mathrm{TM_{0}}}^{\mathrm{tg}}$
and TE mode $\beta_{\mathrm{TE_{0}}}^{\mathrm{tg}}$ are indicated,
respectively, by vertical blue and orange dashed lines. The black
solid line is the light line in vacuum, whereas the grey solid curve
plots the SPP dispersion curve at the silver-silica interface. \label{fig:slabSPP}}
\end{figure}
A metasurface considered in this work is composed of a slab waveguide
and a lattice of nanoparticles (meta-atoms) on its surface. We first
study the properties of the waveguide alone, treating it as a periodic
structure with an empty lattice of meta-atoms \citep{carronResonancesTwodimensionalParticle1986,cherquiPlasmonicSurfaceLattice2019}.
A group of qualitatively similar dispersion relations can then be
found in a metasurface based on the same slab waveguide \citep{carronResonancesTwodimensionalParticle1986,tikhodeevQuasiguidedModesOptical2002}.
Therefore, it is useful to first investigate the modal dispersion
of a slab waveguide and then apply the obtained knowledge to the problem
of the near-field enhancement in a real 2D metasurface.

The waveguide is made of a silica slab on a silver substrate {[}see
Fig.~\ref{fig:slabSPP}(a){]}. The silver substrate can also enhance
the field on the surface by simply reflecting the incident field.
The refractive indices of silver, silica, and air are denoted by $n_{\mathrm{Ag}}$,
$n_{\mathrm{SiO2}}$, and $n_{\mathrm{air}}$, respectively. The dispersive
refractive index of silver is downloaded from the online refractiveindex.info
database \citep{polyanskiyRefractiveindexInfoDatabase2008}, which
uses Johnson and Christy's tabulated data \citep{johnsonOpticalConstantsNoble1972}.
The silica and air regions are treated as having constant refractive
indices of $1.4537$ and $1$, respectively. The waveguide modes are
calculated by solving the following one-dimensional problem in a weak
formulation, using the finite element method (FEM), 
\begin{equation}
\int_{C}dl\left[\frac{1}{\alpha}\frac{\partial v_{z}}{\partial x}\frac{\partial u_{z}}{\partial x}-\left(\frac{\omega}{c}\right)^{2}\gamma v_{z}u_{z}\right]=-\beta^{2}\int_{C}dl\frac{1}{\alpha}v_{z}u_{z},
\end{equation}
where $u_{z}$ is the only unknown transverse component of the mode
as a function of $x$, and $v_{z}$ is the corresponding test function.
For a TM mode, $u_{z}$ denotes the transverse magnetic field and
the domain-wise parameters are $\alpha=\epsilon$ and $\gamma=\mu$.
For a TE mode, $u_{z}$ represents the transverse electric field and
the parameters are $\alpha=\mu$ and $\gamma=\epsilon$. The modal
propagation constant $\beta$ is sought by solving the above eigenvalue
problem for $-\beta^{2}$. We have implemented the weak formulation
calculations using both the Equation-based Module in COMSOL Multiphysics
\citep{comsol} and the open-source software NGSolve \citep{ngsolve},
which generate consistent results. In the calculations, the semi-infinite
air and silver regions are truncated by Perfectly Matched Layers (PMLs),
which are modelled as frequency-dependent, anisotropic absorbing media
\citep{sacksPerfectlyMatchedAnisotropic1995,gedneyIntroductionFinitedifferenceTimedomain2011}.
The outside boundaries of PMLs are those of a perfect electric conductor
for a TE mode and a perfect magnetic conductor for a TM mode.

The field profiles of the fundamental $\mathrm{TM}$ and $\mathrm{TE}$
modes are shown in Figs.~\ref{fig:slabSPP}(b) and \ref{fig:slabSPP}(c),
respectively, for a slab thickness of $t_{\mathrm{slab}}=200$ nm
at the wavelength of $780$ nm. The $\mathrm{TM}$ mode has a dominant
field component $E_{y}$ due to the presence of silver at the bottom
of the waveguide. It is more concentrated inside the slab and resembles
a surface plasmon polariton (SPP) mode at a silver-silica interface
\citep{maierPlasmonicsFundamentalsApplications2007}. However, the
$\mathrm{TE}$ mode has only the $E_{z}$ component that is nearly
negligible at the metal surface and has intensity maxima closer to
the silica-air interface. In addition, the $\mathrm{TM}$ mode has
a shorter modal wavelength than the $\mathrm{TE}$ mode, with the
former being $\lambda_{\mathrm{TM_{0}}}=550$ nm and the latter being
$\lambda_{\mathrm{TE_{0}}}=753$ nm. The modal dispersion curves of
slab waveguides with $t_{\mathrm{slab}}=40$ nm and $t_{\mathrm{slab}}=200$
nm are plotted in Figs.~\ref{fig:slabSPP}(d) and \ref{fig:slabSPP}(e),
respectively. For relatively thin slabs, the fundamental $\mathrm{TM}$
mode approximately follows the dispersion curve of the $\mathrm{SPP}$
on a silver-silica interface at high frequencies, while it approaches
the light line in air at low frequencies. If the slab thickness increases,
more modes emerge and the dispersion curve of the fundamental $\mathrm{TM}$
mode is pushed towards the $\mathrm{SPP}$ curve. The target wavelength
of $780$ nm is shown by the horizontal solid grey line that intersects
with the dispersion curves and determines the modal propagation constants
$\beta_{\mathrm{TM_{0}}}^{\mathrm{tg}}$ and $\beta_{\mathrm{TE_{0}}}^{\mathrm{tg}}$
at this wavelength.

\begin{figure}[tb]
\begin{centering}
\includegraphics{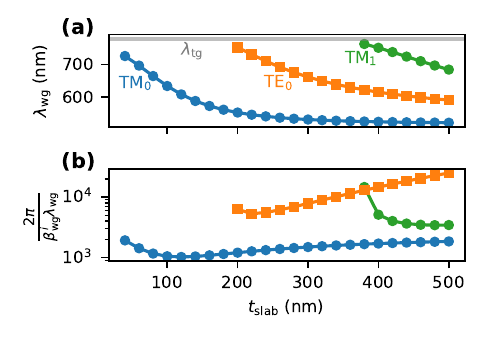}
\par\end{centering}
\caption{(a) Mode wavelengths $\lambda_{\mathrm{wg}}=2\pi/\beta_{\mathrm{wg}}^{r}$
of the slab waveguide; (b) Normalized decay lengths $2\pi/\beta_{\mathrm{wg}}^{i}/\lambda_{\mathrm{wg}}$
for a varying slab thickness. The real and imaginary parts of the
propagation constant are denoted by $\beta_{\mathrm{wg}}^{r}$ and
$\beta_{\mathrm{wg}}^{i}$, respectively. \label{fig:lambda_mode}}
\end{figure}
The wavelengths of the guided modes, $\lambda_{\mathrm{wg}}=2\pi/\beta_{\mathrm{wg}}^{r}$,
with $\beta_{\mathrm{wg}}^{r}$ denoting the real part of the propagation
constant, are shown in Fig.~\ref{fig:lambda_mode}(a) for a varying
slab thickness. In addition, the normalized decay lengths, $2\pi/\beta_{\mathrm{wg}}^{i}/\lambda_{\mathrm{wg}}$
\citep{solymarWavesMetamaterials2009}, are plotted in Fig.~\ref{fig:lambda_mode}(b),
where $\beta_{\mathrm{wg}}^{i}$ is the imaginary part of the propagation
constant. The guided modes can therefore propagate over a distance
of $10^{3}-10^{4}$ modal wavelengths, leading to a long-distance,
highly nonlocal interaction of the meta-atoms.

\section{Bloch modes in plasmonic metasurfaces\label{sec:Bloch_metasurfaces}}

Before discussing the guided Bloch modes and lattice resonances in
our designed metasurfaces, we extend the empty-lattice dispersion
relation of the free-photon approximation \citep{carronResonancesTwodimensionalParticle1986,cherquiPlasmonicSurfaceLattice2019}
to the free-guided-wave approximation. The empty-lattice dispersion
relation is
\begin{equation}
\left|\mathbf{k}_{||}+\mathbf{G}\right|=\beta_{\mathrm{wg}},
\end{equation}
where $\mathbf{k}_{||}=\hat{y}k_{y}+\hat{z}k_{z}$ is the in-plane
Bloch wave vector, $\mathbf{G}=m\frac{2\pi}{\Lambda_{y}}+n\frac{2\pi}{\Lambda_{z}}$
is the reciprocal lattice vector (with the integers $m$ and $n$
denoting the Bloch-mode orders along the $y$ and $z$ directions,
respectively), and $\beta_{\mathrm{wg}}$ is the propagation constant
of the slab mode. The slab waveguide is homogeneous and extends to
infinity in the transverse plane. However, the fictitious periodicity
causes the band folding \citep{carronResonancesTwodimensionalParticle1986,cherquiPlasmonicSurfaceLattice2019,tikhodeevQuasiguidedModesOptical2002,yannopapasOpticalExcitationCoupled2004}.
If we choose the periods to be equal to the guided-mode wavelengths,
the empty-lattice Bloch modes appear to be formed at the $\Gamma$-point
with $k_{y}=k_{z}=0$, which can be used for designing the Bloch modes
in a periodic metasurface to be excited at normal incidence \citep{yannopapasOpticalExcitationCoupled2004}.

\begin{figure}[tb]
\begin{centering}
\includegraphics[scale=0.73]{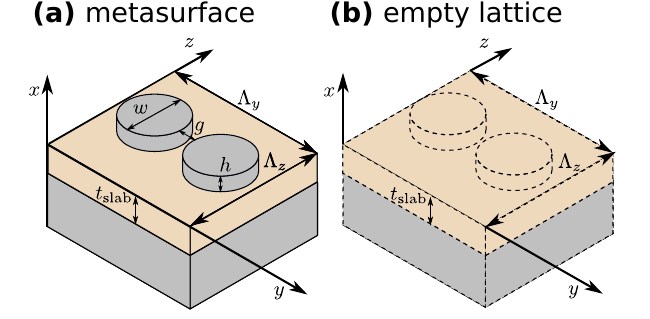}
\par\end{centering}
\begin{centering}
\includegraphics{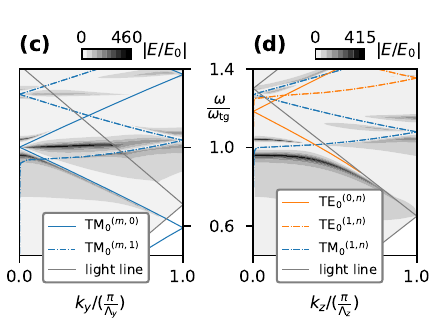}
\par\end{centering}
\caption{Schematic diagrams of (a) a metal-dielectric metasurface and (b) the
corresponding empty lattice. The dimer consists of two identical nanodiscs
with a diameter of $w$ and a height of $h$, which are separated
by a gap of $g$. The two in-plane periods are $\Lambda_{y}$ and
$\Lambda_{z}$. In the empty lattice, the dashed outline implies that
the nanodiscs are actually removed. Note that the dimers in our calculations
are about two times smaller than in the pictures. The parameter values
used in the calculations are $\Lambda_{y}=550$ nm, $\Lambda_{z}=600$
nm, and $t_{\mathrm{slab}}=200$ nm. The near-field enhancement in
the metasurface and the empty-lattice dispersion curves are shown
in (c) and (d).\label{fig:metasurface_vs_emptylattice}}
\end{figure}
In a metasurface containing silver meta-atoms, the meta-atomic LSPRs
are coupled to the empty-lattice Bloch modes, resulting in modified
Bloch modes \citep{cherquiPlasmonicSurfaceLattice2019}. The unit
cell of the metasurface is shown in Fig.~\ref{fig:metasurface_vs_emptylattice}(a).
It contains a meta-atom in the form of a silver-disc dimer with dimensions
$w=120$ nm, $h=20$ nm, and $g=6$ nm. The values of other parameters
of the structure are $t_{\mathrm{slab}}=200$ nm, $\Lambda_{y}=550$
nm, and $\Lambda_{z}=600$ nm. At $\lambda=780$ nm, the period $\Lambda_{y}$
is equal to the mode wavelength $\lambda_{\mathrm{TM_{0}}}$, but
the period $\Lambda_{z}$ is shorter than $\lambda_{\mathrm{TE_{0}}}$.
Note that plasmonic dimers are among the most efficient standard nanoparticles
used to locally enhance optical fields, and dimers composed of nanodiscs
rather than particles with sharp corners are relatively easy to nanofabricate
and treat numerically.

The near-field enhancement spectra in the nonlocal metasurface are
calculated using the frequency-domain FEM\textcolor{red}{{} }of the
Wave Optics Module of COMSOL Multiphysics \citep{comsol}. The background
excitation field is either a $p$- or $s$-polarized plane wave in
the scattered-field formulation. A $p$-polarized incident plane wave
(with the electric field lying in the $xy$-plane) couples effectively
to the TM Bloch modes of the structure, yielding Fano-type \citep{khlopinLatticeModesPlasmonic2017}
asymmetric line shapes in the near-field enhancement spectra. This
is revealed in Fig.~\ref{fig:metasurface_vs_emptylattice}(c), where
high (in black) and low (in white) near-field enhancement factors
appear near the empty-lattice dispersion curves. A band gap showing
low near-field enhancement is formed around the normalized frequency
$\omega/\omega_{\mathrm{tg}}=1$ at $k_{y}=0$, as a result of coupling
of the LSPR with the empty-lattice Bloch modes $\mathrm{TM_{0}^{(+1,0)}}$
and $\mathrm{TM_{0}^{(-1,0)}}$ that counter-propagate along axis
$y$. For higher-order empty-lattice TM modes with order $n=1$ (the
blue dashed curve), another band gap is formed near the normalized
frequency $\omega/\omega_{\mathrm{tg}}=1.04$ at $k_{y}=\pi/\Lambda_{y}$.

If light is incident in the $zx$-plane and $s$-polarized, it couples
effectively to the Bloch modes propagating in the $z$-direction.
The excited LSPRs in the dimers still provide a high gap enhancement.
In Fig.~\ref{fig:metasurface_vs_emptylattice}(d), one can observe
the near-field enhancement spectra shown in band gaps near $\omega/\omega_{\mathrm{tg}}=1.0$
and $1.08$ due to the interaction of LSPRs with the $\mathrm{TM}_{0}^{(1,n)}$
modes at $k_{z}=0$ and $k_{z}=\pi/\Lambda_{z}$, respectively. The
interaction leads to high near-field enhancement factors (see the
dark regions in the figure). Additional empty-lattice TE and TM modes
do not significantly affect the near-field enhancement and they are
not shown in the figure.

\section{Near-field enhancement in a planar metasurface\label{sec:nf_metasurface}}

\begin{figure}[tb]
\begin{centering}
\includegraphics[scale=0.9]{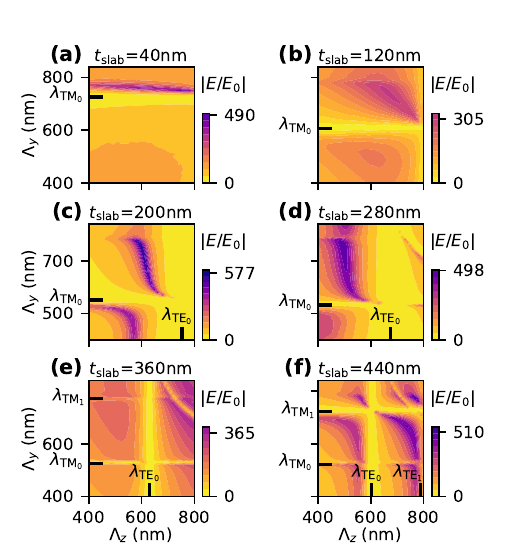}
\par\end{centering}
\caption{Near-field enhancement factor $|E/E_{0}|$ as a function of periods
$\Lambda_{y}$ and $\Lambda_{z}$ for several slab thicknesses $t_{\mathrm{slab}}$
obtained at the fixed wavelength $\lambda_{\mathrm{tg}}=780$ nm.
Thick black ticks with labels $\lambda_{\mathrm{TM}_{0}}$, $\lambda_{\mathrm{TM}_{1}}$,
and $\lambda_{\mathrm{TE}_{0}}$ mark the guided-mode wavelengths
in the slab waveguide of the corresponding thickness.\label{fig:nearfield_metasurface}}
\end{figure}

We can now maximize the near-field enhancement by optimizing the periods
$\Lambda_{y}$ and $\Lambda_{z}$, as well as the slab thickness $t_{\mathrm{slab}}$.
As shown in Fig.~\ref{fig:nearfield_metasurface}(a-f), both the
periods affect the near-field enhancement, and the fingerprints of
band gaps are recognized as straight stripes of low near-field enhancement
near the periods equal to the guided-mode wavelengths (indicated by
thick black ticks with labels $\lambda_{\mathrm{TM}_{0}}$, $\lambda_{\mathrm{TM}_{1}}$,
and $\lambda_{\mathrm{TE}_{0}}$). The Bloch modes formed above or
below the guided-mode wavelengths contribute the most to the near-field
enhancement in the gaps of the nanoparticles. A maximum field enhancement,
$|E/E_{0}|=577$, is achieved at periods $\Lambda_{z}=595$ nm and
$\Lambda_{y}=760$ nm for the slab thickness $t_{\mathrm{slab}}=200$
nm. Such a configuration allows for a SERS enhancement factor, $\eta_{\mathrm{SERS}}$,
of the order of $10^{11}$ at a gap size of $6$ nm. At smaller gaps,
the enhancement factor is larger, but we consider 6 nm as a reasonable
gap size that can be obtained in nanofabricated samples.

\begin{figure}[tb]
\begin{centering}
\includegraphics{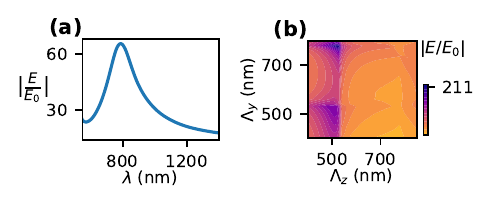}
\par\end{centering}
\caption{Near-field enhancement factors of (a) a single dimer as a function
of wavelength $\lambda$ and (b) a lattice of dimers on top of a semi-infinite
silica as a function of periods $\Lambda_{y}$ and $\Lambda_{z}$.\label{fig:nearfield_refSamples}}
\end{figure}
It is interesting to compare the near-field enhancement in the optimized
metasurface to that of a single dimer and a lattice of dimers on a
thick silica substrate. For an isolated single dimer on a thick substrate,
the near field can be resonantly enhanced by a factor of $65$ at\textbf{
$\lambda=780$} nm, as shown in Fig.~\ref{fig:nearfield_refSamples}(a).
If such dimers are arranged periodically in a lattice on the same
substrate, an extra factor of $3$ can be obtained for the near-field
enhancement {[}see Fig.~\ref{fig:nearfield_refSamples}(b){]}. In
our design, with a silver mirror positioned at a distance of $200$
nm below the surface, an overall near-field enhancement of $577$
is obtained. Hence, arranging the particles in a lattice on a slab
waveguide resulted in about $80$ times higher local-field intensity
in the gaps compared to the case of isolated dimers on a thick substrate,
providing an extra SERS enhancement of more than $6000$ (calculated
as a squared intensity enhancement). The presence of the waveguide
additionally increases the local-field intensity by more than $7$
times, corresponding to a $50$-fold increase of $\eta_{\mathrm{SERS}}$.
It is well known that naked metal nanostructures of typical SERS substrates
are not suitable for multiple use because of their fast and irreversible
degradation caused, e.g., by humidity variations, surface oxidation,
and contamination. The structures could be protected with a few-nm-thick
dielectric coating, but this usually lowers $\eta_{\mathrm{SERS}}$
by orders of magnitude. The extra enhancement considered in this work
can compensate for this decrease of SERS signal and allow making the
substrates reusable.

\begin{figure*}[t]
\begin{centering}
\includegraphics[scale=0.85]{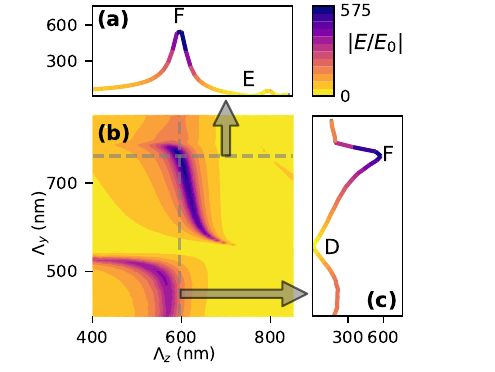}\includegraphics[scale=0.85]{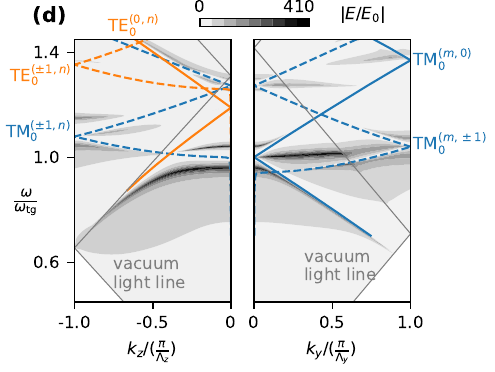}
\par\end{centering}
\begin{centering}
\includegraphics[scale=0.85]{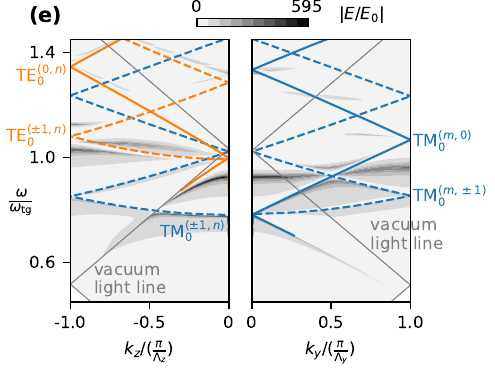}\includegraphics[scale=0.85]{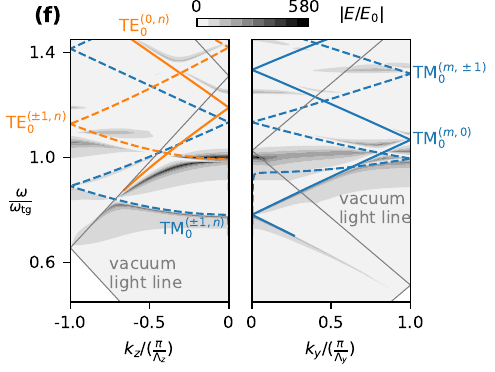}
\par\end{centering}
\caption{Near-field enhancement spectra for $t_{\mathrm{slab}}=200$ nm and
(a) a fixed period $\Lambda_{y}=760$ nm, (b) varying periods $\Lambda_{y}$
and $\Lambda_{z}$, and (c) a fixed period $\Lambda_{z}=595$ nm.
For three typical configurations at points D, E, and F, the near-field
enhancement spectra as a function of the incidence angles along the
$y$ and $z$ axes are shown in (d), (e), and (f), respectively. Blue
and orange lines indicate the empty-lattice Bloch modes $\mathrm{TM}_{0}^{(m,n)}$
and $\mathrm{TE}_{0}^{(m,n)}$, respectively. In both colours, solid
lines indicate the zero-order modes, whereas the dashed lines stand
for the $1$st-order modes.\label{fig:typical_configurations}}
\end{figure*}

We further explore the near-field enhancement spectra of three designed
metasurfaces with $t_{\mathrm{slab}}=200$ nm, marked by points D,
E, and F in Fig.~\ref{fig:typical_configurations}(a)-(c). For the
metasurface at point D, the periods are $\Lambda_{y}=550$ nm and
$\Lambda_{z}=595$ nm, and the near-field enhancement is very low
due to the band gap near the normalized frequency $\omega/\omega_{\mathrm{tg}}=1$
at $k_{y}=0$ {[}see Fig.~\ref{fig:typical_configurations}(d){]}.
This band gap is formed because the period $\Lambda_{y}$ matches
with the modal wavelength $\lambda_{\mathrm{TM}_{0}}=550$ nm of the
fundamental $\mathrm{TM}$ mode propagating along the $y$-axis {[}see
Fig.~\ref{fig:lambda_mode}(a){]}.

At point E, the periods are $\Lambda_{y}=760$ nm and $\Lambda_{z}=755$
nm, with the latter being close to the modal wavelength $\lambda_{\mathrm{TE}_{0}}=753$
nm of the fundamental $\mathrm{TE}$ mode propagating along the $z$-axis
{[}see Fig.~\ref{fig:lambda_mode}(a){]}. This results in a band
gap near the normalized frequency $\omega/\omega_{\mathrm{tg}}=1$
at $k_{z}=0$ {[}see Fig.~\ref{fig:typical_configurations}(e){]},
which makes the near-field enhancement low.

The maximum near-field enhancement is obtained for the metasurface
at point F with a pair of periods $\Lambda_{y}=760$ nm and $\Lambda_{z}=595$
nm. As shown in Fig.~\ref{fig:typical_configurations}(f), the upper
band of the guided Bloch mode is formed at the normalized frequency
$\omega/\omega_{\mathrm{tg}}=1$ at $k_{y}=0$, because $\Lambda_{y}$
is larger than the modal wavelength $\lambda_{\mathrm{TM}_{0}}$.
On the other hand, the lower band of the guided Bloch mode is formed
at the normalized frequency $\omega/\omega_{\mathrm{tg}}=1$ at $k_{z}=0$,
as $\Lambda_{z}$ is smaller than the modal wavelength $\lambda_{\mathrm{TE}_{0}}$.
The upper and lower bands of the guided Bloch modes counter-propagating
along the $y$ and $z$ axes jointly enhance the near-field in the
gaps of the dimers.

\section{Conclusions\label{sec:Conclusions}}

We have shown that, with the aid of guided Bloch modes in a designed
nonlocal optical metasurface, the local field intensity can be considerably
increased in addition to the enhancement by isolated plasmonic nanoparticles.
The additional enhancement mechanism relies on the coupling of the
incident light to the LSPRs of individual nanoparticles and to the
Bloch modes guided in a metal-dielectric slab waveguide beneath the
particles. The coupling takes place at the second stop band edge for
normally incident light. The maximum enhancement is obtained by optimizing
the two periods and the thickness of the waveguide. The approach clearly
outperforms the lattice-resonance-based enhancement. For plasmonic
nanodimers with a gap size of $6$ nm, we obtained a SERS enhancement
of $10^{11}$. By reducing the gap, the enhancement factor can be
increased by further orders of magnitude. In addition, using dimers
with sharp edges, such as bowtie antennae, would further enhance the
local field intensity. We expect that the proposed mechanism of additional
near-field enhancement will find applications not only in SERS, but
also in fluorescence- and scattering-based plasmonic sensing, light-energy
harvesting components, and nonlinear optics.

\ack{}{}

We acknowledge the Academy of Finland Flagship Programme Photonics
Research and Innovation (PREIN; grant No.~320167). We thank Dr.~R.~Kolkowski
for fruitful discussions and advice on theoretical simulations of
metasurfaces and slab waveguide modes. We also acknowledge the computational
resources provided by the Aalto Science-IT project.

\bibliographystyle{unsrturl}
\phantomsection\addcontentsline{toc}{section}{\refname}\bibliography{references}

\end{document}